# Fair Stateless Aggregate Traffic Marking using Active Queue Management Techniques

Abhimanyu Das,    Debojyoti Dutta,    Ahmed Helmy

*Abstract*—In heterogeneous networks such as today's Internet, the differentiated services architecture promises to provide QoS guarantees through scalable service differentiation. Traffic marking is an important component of this framework. In this paper, we propose two new aggregate markers that are stateless, scalable and fair. We leverage stateless Active Queue Management (AQM) algorithms to enable fair and efficient token distribution among individual flows of an aggregate. The first marker, Probabilistic Aggregate Marker (**PAM**), uses the Token Bucket burst size to probabilistically mark incoming packets to ensure TCP-friendly and proportionally fair marking. The second marker, Stateless Aggregate Fair Marker (**F-SAM**) approximates fair queueing techniques to isolate flows while marking packets of the aggregate. It distributes tokens evenly among the flows without maintaining per-flow state. Our simulation results show that our marking strategies show upto 30% improvement over other commonly used markers while marking flow aggregates. These improvements are in terms of better average throughput and fairness indices, in scenarios containing heterogeneous traffic consisting of TCP (both long lived elephants and short lived mice) and misbehaving UDP flows. As a bonus, F-SAM helps the mice to win the war against elephants.

## I. Introduction

The Differentiated Service architecture *(Diffserv)* [1] [2] addresses the issue of providing statisitical QoS guarantees within IP by classifying and aggregating flows into different classes, and providing service differentiation among the traffic aggregates. The Diffserv approach does not suffer from the scalability problems of IntServ [3] due to its stateless nature. Diffserv uses fine grained, per-flow marking at the network edges to classify flows into the various classes, and, then, applies coarser per-class service differentiation (Per-Hop Behavior) at the network core. There are two main types of Per-hop Behaviors(PHB) currently defined, the Expedited Forwarding(EF) PHB, and the Assured Forwarding(AF) PHB.

In this paper we focus on packet marking for an Assured Forwarding PHB domain [4]. Traffic is marked at the edge into different drop-priority classes, according to a service profile. The network core uses simple AQM techniques to provide preferential packet dropping among the classes. For simplicity, we assume that there are two AF classes, IN and OUT, and, the core routers use RIO (RED with In/Out). Before Diffserv flows from different edge domains enter the network core, they need to be marked as IN or OUT at their respective edge networks based on a traffic service profile on the basis of a service level agreements (SLA) between the two domains (negotiated aproiri) This SLA is typically modeled as a token bucket for each diffserv class (in our case, a traffic specification for IN packets).

Since the SLA at an edge domain applies to the egress traffic, we need to perform traffic marking on the aggregated traffic at the edge. Using simple, popular token bucket based markers at the edge however, leaves no scope to distinguish between flows within an aggregate. In particular, there is no way to prevent excessive token allocation to misbehaving, non-responsive flows. Further, even within well-behaved TCP flows of an aggregate, there is no provision for preventing long TCP flows from squeezing out short duration web traffic. Token Bucket markers tend to aggravate bursty behavior and increase packet drops, and this can severely affect TCP performance.

Thus there is a need to ensure to develop a marker that allocates tokens fairly between responsive(TCP) and non responsive(UDP) flows and between long and short TCP flows. We also need to ensure that markers be *TCP friendly* and scalable (as the number of flows increases).

Note that most of the aforementioned issues in marking are also common to those in the design of Active Queue Management [5] schemes. Well established work in active queue management such as RED [6], CHOKE [7] and CSFQ [8] have been known to provide certain degrees of fairness to flows and/or better throughput for TCP flows by active probabilistic dropping of packets from the queue, without maintaining per-flow state. We draw an analogy between Diffserv packet marking and queue management to formulate our methods. In our work, we utilize these AQM mechanisms to propose two alternate methods of edge aggregate marking in diffserv edge routers for allocating tokens to an aggregate of incoming flows. We show through extensive simulations that our methods achieve fairness in token allocation between flows without maintaining per-flow state. In addition, both the methods are probabilistic in nature. This helps to decrease bursts of downgraded packets belonging to a single flow, thus making the marked flows more TCP-friendly [9]. Finally we also show that one of our two markers, F-SAM, can also provide a marked improvement in the performance of short TCP flows in a loaded network scenario, and prevent short web flows from getting unfairly squeezed out during bandwidth contention with longer flows.

We are aware of marking techniques that individually address some of the issues with aggregate edge marking which we highlighted earlier. But none of them address all these aforementioned issues simultaneously, or they are complex and need per-flow processing. We conjecture that applying ideas from AQM techniques can be promising in the diffserv packet marking area. As we see in our simulation results, it turns out to be efficient and scalable.

The rest of this document is outlined as follows. Section II presents related work on markers. A brief background on traffic marking is given in Section III and our first scheme, PAM, is describes in Section IV. Section V details our second scheme, F-SAM. Our simulation results are discussed in Section VI. Section VII outlines future work and Section VIII concludes.

The authors are with University of Southern California, Los Angeles, USA. E-mail: abhimand@cs.usc.edu, ddutta@isi.edu and helmy@usc.edu

## II. RELATED WORK

Packet marking is a well visited topic. Many current markers perform sophisticated per-flow marking. Many current aggregate traffic markers mark individual packets of an aggregate using simple tokenbucket-like schemes without looking at fair token distribution among individual flows; a requirement we believe to be essential to providing fairness in QoS based on the diffserv architecture. Also most traffic markers do not look at TCP-friendly nature of their marking, or the treatment of short web flows in the presence of long TCP flows.

There has been some work in aggregate marking, which do not use per-flow state. They typically rely of metering the average rates and comparing them against a token traffic specification [10], [11]. However these markers do not look at fairness issues within an aggregate.

A sophisticated aggregated marker can be found in Fang. et. al. [12]. It does probabilistic marking based on the average aggregate rates, and in this respect is slightly similar to one of our proposed markers. However instead of a RED-like transfer function on the burst size, as in our case, they use the average incoming rate and not the burst size. Thus they do not take care to remove burstiness of TCP flows. Further they too do not address the bias against short-TCP flows while marking in a congested network.

Maintaining individual, per-flow state can, of course solve our fair aggregate marking problem. However, these schemes are not scalable or even very feasible when the number of flows entering a marker is varying and dynamic. Yeom and Reddy [13] also looks at fair aggregate marking issues, but they assume that the individual flows within the aggregate have already been pre-allocated individual contract rate from the aggregate marker tokens. Also their algorithm is per flow based, and entails calculating individual flow throughput information using a TCP throughput model which requires Round Trip Time and PacketSize information for each flow. Besides, it does not apply to non-TCP flows.

In [14], the authors look specifically at the issue of sharing excess network bandwidth among traffic aggregates. However, they do not investigate the problem of fairly marking individual flows within an aggregate, which is the problem we address in this study. So their aggregate marker looks at inter-aggregate fairness as opposed to intra-aggregate fairness in our case.

In [9] the authors aim to give preferential treatment to short-lived TCP flows in an aggregate and perform a max-min fairness on the remaining TCP flows. But they assume that their marker has state about all the component TCP flows in the aggregate, which might not be practical or scalable. Furthermore they only address TCP flows which might not be adequate in todays heterogeneous networks.

In [15] the authors propose using an FRED algorithm to create a fair aggregate traffic marker. However this require per-active flow computation at the marker and is also more complex to implement than either of our markers. Also, they do not address the issues in fairness between long-lived and short-lived TCP flows.

We are not aware of any previous work that aims to solve the problem of fair token distribution among heterogeneous flows within an aggregate without maintaining per-flow state, within the diffserv framework. Besides, we are not aware of any marking techniques based on AQM techniques such as RED [6] and CSFQ [8].

## III. TRAFFIC MARKING

Traffic marking is an essential component of the diffserv architecture. The traffic marker looks at the incoming packets and compares it with a given traffic profile (for example, a token bucket(TB) characterization of a SLA between two domains). In-profile packets are marked with an identifier that indicates a higher priority. If packets are out of profile, it is *marked* with a lower priority.

The key issue here is an efficient way of distributing the allocated tokens (specified in the SLA traffic specification of IN packets) among individual flows of the edge network entering the marker. One way of doing this is to apriori divide up the aggregate SLA token distribution into smaller token traffic specifications for individual flows (or groups of flows), generated by hosts in the domain, according to some administrative policy within the edge domain. In this case, the edge marker would have knowledge of individual token bucket specifications for every incoming flow. Thus the issue of fairness in distributing tokens among the flows(or groups of flows) in an aggregate, is now handled by policy-driven distribution of tokens among flows at the marker. Clearly, this approach, while trivially solving all fairness issues in token distribution, is not very practical or scalable. The number and type of flows generated in a domain can be dynamic and varying, and so it is not easy to micro-manage token distribution to individual flows.

Note that this problem of aggregate marking is quite similar to the issue of queue management and scheduling, which consists of a queue with a finite set of buffers for storing incoming packets and sending them out a limited bandwidth outgoing link. The queue receives packets from various flows and has to decide which packets from the incoming aggregate to buffer and which to drop. Fairness and efficiency in throughput allocation among individual flows being served by the queue is a key issue here.

We can therefore view a token bucket specification as a queue with the TB Burst Size as the queue size and the token arrival rate as the queue's link bandwidth. Marking an arriving packet as IN is similar to queueing a packet for transmission, and marking it as OUT is equivalent to dropping an incoming packet from the queue. Both the average queue size and the average token bucket size give an indication of the congestion level at the queue/TB. However a small average queue size is equivalent to a large average tokenbucket size and vice versa. The problem of fair token distribution at a marker (using a Token Bucket traffic specification) among packets of various incoming flows, can be viewed as being equivalent to efficient buffer management and scheduling of incoming packets at a queue to provide fair bandwidth distribution among the flows arriving at the queue.

Active Queue Management techniques such as Random Early Detection(RED) [6] proactively drop packets before a queue overflow event, and allow for higher throughput and better fairness by reducing synchronization effects, bursty packet losses and biases against low bandwidth-bursty TCP flows. In addition, queue management techniques like Core Stateless Fair Queueing(CSFQ) [8], Flow Random Early Detection(FRED) [16],

CHOKE [7] and Fair Queueing [17] try to distribute link bandwidth at the queue fairly among all flows of the aggregate.

We therefore try to apply two sample AQM techniques - CSFQ and RED to aggregate diffserv packet marking using token bucket specifications We observe that these techniques are superior to current markers in terms of throughput and fairness.

In the next section, we introduce our two stateless AQM based marking schemes. The first one is based on RED and the second one is based on CSFQ.

## IV. PROBABILISTIC AGGREGATE MARKER

A simple aggregate token bucket marker (which includes single rate TCM and a two rate TCM) behaves like a drop-tail queue. As long as the token bucket has tokens, incoming packets will be marked as IN and once the token bucket is empty, packets arriving at the marker will be marked as OUT. Many of the problems related to drop-tail queues therefore also apply to these types of markers. They are biased against low-bandwidth and bursty flows and can cause consecutive packets of a single TCP flow to be marked as OUT. They are not therefore very fair or "TCP-friendly" [9].

Our Probabilistic Aggregate Marker (PAM) is based on *RED (Random Early Detection)* [6]. Intuitively, the idea is as follows. The aggregate traffic will try to consume tokens from the token bucket at a certain rate. We maintain an exponentially weighed moving average (EWMA) of the number of tokens in the token bucket. On every incoming packet, we look at this average token bucket size. If this bucket size falls below a certain threshold $min_{th}$, all packets are to be marked with a lower priority. If the bucket size varies between $min_{th}$ and $max_{th}$, we mark the packet as OUT based on a probability function that depends on the instantaneous size of the token bucket. If the token bucket size exceeds $max_{th}$, we mark the packet as IN. More formally, our probability function of marking as OUT $P(x)$ can be written as follows.

$$P(x) = \begin{cases} 1 & : x < min_{th} \\ \frac{P_{max} - P_{min}}{max_{th} - min_{th}} \times (max_{th} - x) & : min_{th} < x < max_{th} \\ 0 & : x > max_{th} \end{cases}$$
(1)

where $x$ is the average size of the token bucket. Our probability function for marking a packet as OUT, therefore a modification of the RED probability function for accepting a packet into the queue.

The rationale behind using a probabilistic RED-like approach is that we want to apply aggregate stateless marking, maintain more TCP-friendly and equitable marking among the flows of the aggregate, without trying to explicitly identifying individual flows. Like RED, this marking scheme ensures proportional marking, since every packet is marked as OUT with a certain probability (based on the average token bucket size) once the token bucket starts getting depleted. Hence, the flow that pumps more traffic into the edge will, on the average, have a greater number of OUT packets, and hence will have more of its packets dropped if the core is congested. We believe that this scheme will be fairer to TCP flows in the presence of a misbehaving non-responsive flow than simple token bucket markers, since it remarks packets of a flow proportionally to the number of packets being sent by the flow. Thus, a high-bandwidth misbehaving flow will get a lower fraction of the IN tokens than in a simple tokenbucket marker case.

We claim that performance of TCP flows in a congested network using PAM will be better than that using a TB marker. Also in case of a misbehaving UDP flow, PAM will ensure better throughput to TCP flows than TB. We test our claims using simulation and present the results in Section VI. One major advantage of this approach is that it is very simple to implement and does not keep per-flow state.

## V. FAIR STATELESS AGGREGATE MARKER

In this section, we propose a very efficient, stateless aggregate fair marker *F-SAM*. This uses a max-min fairness criteria [17], fair queueing to performs probabilistic marking and ensure TCP-friendliness. The idea here is to apply approximate fair queuing to the token bucket, while distributing the tokens among the packets of the flows in the aggregate, but without maintaining any per flow queue or state.

Stoica et al. [8] propose a Core Stateless Fair Queueing algorithm to approximate max-min fairness while buffering incoming packets at a queue, by using a probabilistic dropping function based on the average rate of a flow to which the packet belongs. This rate information, instead of being calculated at the queue using per-flow techniques is calculated near the source of the flow and inserted in every packet header. We adapt the core stateless fair queuing to aggregate packet marking. We maintain the rate information of a flow in the packet header itself, and use this to calculate a token allocate probability, on a per-packet basis in the edge marker. The queue, output link speed and drop probability in [8] a re replaced by the token bucket, token-bucket rate and $(1 - P[token\ allocation])$. Packets which get through the probabilistic dropping function based on their rate information, and manage to get tokens from the token bucket are marked as IN, while others are marked as OUT.

In F-SAM, the rate information in each packet header is calculated and filled by the ingress node when the flow enters the domain. Since each ingress node is responsible for maintaining the rate of only the flow that enters through it, there is no scalability issue involved in this per-flow rate calculation. At the egress, the edge marker needs to calculate the fair rate, $f$, allocated to the flows, and then calculates the token allocation probability (marking probability) of a packet as $min(1, \frac{f}{r})$, where $r$ is the rate of the corresponding flow. So the expected token allocation rate to a packet belonging to a flow of rate $r$, is then $min(r, f)$, which corresponds to the max-min fairness of fair-queuing. The aggregate token allocation rate $R(f)$ will then be $\sum_i min(r_i, f)$. As in [8], the fair-rate $f$, is calculated by an iterative method based on the average aggregate arrival rate of packets $A$, estimate of the aggregate token allocation rate of packets $R'$, and the token bucket rate $C$.

The aggregate arrival rate and token allocation rate are exponentially averaged at the egress marker. Note that the fair-rate calculation algorithm does not need any other per-flow measurements. Due to lack of space, we do not go into the specific algorithm details (see [8] or [18] for more details.)

Thus, this scheme will ensure that tokens are allocated among the various flows fairly, by approximating a max-min fairness token allocation strategy, based on the rate information in the packet headers.

We claim that F-SAM ensures fair token allocation. It is much fairer than any of the standard token bucket schemes and even our PAM marker of the previous section. For example, an aggregate F-SAM marker marking packets containing multiple TCP flows and UDP flows would try and allocate close to an equal number of tokens to each individual flow. This ensures a fairer division of the total throughput among all the flows, both TCP and UDP. Note that even though the token distribution of IN packets to all the TCP and UDP flows is equitable, this will not translate to an exactly equal division of throughput at the network core among a TCP and a UDP flow, if the core only implements simple RIO queueing as in our simulations. This is because the loss of any packet of a TCP flow would result in its cutting back on its sending rate, while it does not affect the sending rate of the UDP flow. However, preventing large UDP flows from getting much more IN tokens than TCP flows would significantly increase the throughput obtained by the TCP flows in presence of misbehaving traffic. Since this is a probabilistic marker, it will be more TCP-friendly in terms of removing a bias against bursty flows. Thus, even in the case of all-TCP aggregate traffic, the throughput and fairness obtained by individual TCP flows will be much higher than simple token bucket markers.

Another important property of F-SAM is that it can eliminate bias against short TCP flows in aggregate traffic consisting of both long and short TCP flows. This is an important property and has been the subject of many papers. The reason for this inherent bias in most queueing mechanisms is that short flows do not get enough time to ramp up their sending rate and RTT estimation, and any packet loss from short flows can be harmful. [9] and [19] look at ways to remove this bias against short TCP flows in the marking and queueing domains. However these papers use per-flow mechanisms to reduce this bias, while F-SAM is a completely stateless marker which reduces this bias by the very nature of its probabilistic marking based on average rate of each flow. To provide a theoretical explanation of how F-SAM preferentially treats short flows, one needs to keep in mind that in the CSFQ algorithm the dropping probability is inversely proportional to the flow rate specified in the packet header. Now, if the rate estimation is an Exponentially Weighted Moving Average, then for short flows, the rate estimation calculation is lower than the actual sending rate of the flow, since the initial rate of the short flow is low and the EWMA does not have time to ramp up before the flow ends. So since packets belonging to short flows arriving at the F-SAM marker have a lower rate in their header than their actual sending rate, the short flows gets a slightly bigger proportion of IN packets than what they are actually entitled to in a fair queueing scheme. This translates into better throughput for the short flows, as we shall see in our results.

We will highlight all these claims in our results section, by comparing the throughput and fairness index of marked traffic from different markers consisting of both a mix of long and short flow TCP traffic and of UDP and TCP flows.

## VI. RESULTS

In this section we evaluate the use of AQM strategies to design better traffic markers. In particular, we evaluate the two markers presented in this paper. We investigate the validity of our claims in the previous sections using packet level network simulation. First, we study scenarios with few flows to gain insight into the marking schemes. Then we investigate scenarios for many flows to compare our aggregate markers against the standard aggregate markers such as TB and TSW2CM.

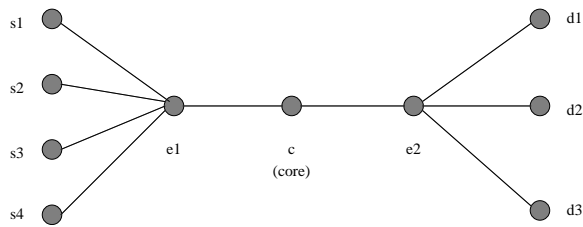

Fig. 1. Simulation scenario: A simple core with two edges. The $s_i$ inject traffic of different classes into the network while the $d_i$s are the sinks.

### A. Experimental Setup

We simulated the design of our markers with the *ns-2* [20] network simulator [1]. The topology used in our experiments is depicted in Figure 1. The source nodes are $s_i$ and the destination or sink nodes are $d_i$. The source nodes are connected to the edge node $e1$ which inject traffic into the network core, $c$. The core is connected to the edge $2e$ which is further connected to the sink nodes $d_i$s. The source node $s_3$ is used to generate background traffic in form of many TCP flows carrying bulk traffic. This background traffic is marked with a separate DSCP (e.g., 20) and is absorbed at $d_2$.

### B. Validating the marking process

| | TCP Flow 1 | | TCP Flow 2 | | Misbehaving UDP Flow | |
|---|---|---|---|---|---|---|
| Marker | IN pkts | OUT pkts | IN pkts | OUT pkts | IN pkts | OUT pkts |
| TokenBucket | 67 | 1261 | 76 | 1267 | 2423 | 7215 |
| PAM | 236 | 935 | 378 | 1456 | 2011 | 7607 |
| F-SAM | 740 | 882 | 748 | 835 | 1042 | 8254 |

Fig. 2. Detailed Packet Marking results with 2 TCP flows and 1 misbehaving UDP flow. For every marker, we give the number of IN and OUT packets obtained by each flow at the marker. The UDP flow has a rate of 2Mb/s.

| | UDP Flow 1 (1Mb/s) | | UDP Flow 2 (4Mb/s) | | UDP Flow 3 (2Mb/s) | |
|---|---|---|---|---|---|---|
| Marker | IN pkts | OUT pkts | IN pkts | OUT pkts | IN pkts | OUT pkts |
| TokenBucket | 13 | 748 | 805 | 2869 | 387 | 1593 |
| PAM | 245 | 695 | 695 | 2883 | 387 | 1593 |
| F-SAM | 497 | 620 | 413 | 2952 | 456 | 1571 |

Fig. 3. Detailed Packet Marking Results with three UDP flows of varying bandwidths. For every marker, we give the number of IN and OUT packets obtained by each flow at the marker.

In this subsection, we study the behavior of our PAM and F-SAM markers with respect to the distribution of tokens among individual flows in an aggregate. We conduct simulations for a mix of UDP and TCP flows and for a mix of UDP flows of varying bandwidth. We verify that the proportion of packets marked for each flow by an aggregate marker conforms to our marker design. This result helps us to estimate the end-to-end performance results. Due to lack of space, we present a very brief summary. A detailed description can be found in [18].

Consider the table in Figure 2. This experiment had 2 TCP flows multiplexed with 1 misbehaving UDP flow(2Mb/s) injected into the core. For our markers, we used a committed information rate (CIR) of 500kb/s and a burst size of 100kb and ran the simulation for 40 seconds in virtual time. We use background TCP traffic in the topology to prevent flow synchronization. For each of the Tokenbucket, PAM and F-SAM markers, we calculate the number of IN and OUT packets marked for each flow. In the token bucket case, we see that the misbehaving UDP flow squeezes the elastic TCP connections in terms of obtaining IN tokens from the marker. However, a comparison of the PAM results with the TB marker results shows that as we increase the sending rate, PAM marks IN packets proportionately i.e. the misbehaving flow gets penalized more (in terms of getting lesser IN packets than in the TB case). In other words, with PAM the ratio $\frac{IN\ packets}{total\ packets\ transmitted}$ is the same for all flows.

However we must note that in a PAM marker, a very large misbehaving flow would take away some fraction of tokens from other conforming flows, and consequently affect marking of all other flows in an aggregate. F-SAM on the other hand, with its max-min fairness approximation, shows a fairer token allocation. In the F-SAM case, we see that all the flows get approximately similar share of tokens from the marker, in spite of the large misbehaving UDP flow (which misappropriated most of the tokens from the TCP flows in the case of the token bucket marker). Thus we see, that the F-SAM marker effectively segregates flows from each other, and does not let misbehaving flows take away tokens from other well-behaved flows.

In the next experiment, we consider only three UDP flows carrying CBR traffic, with varying transmission rates (1Mb/s, 4Mb/s, and 2Mb/s). Looking at Figure 3, we again see a similar improvement of PAM and F-SAM over a TokenBucket Marker in the fairness of token allocation among the three flows. With PAM, the number of tokens allocated to each flow( number of IN packets) is proportional to the total number of packets carried by the flow, thus ensuring proportional fairness.

With F-SAM, the number of tokens allocated to each flow is nearly the same, in spite of widely varying individual flows rates. Comparing with the token bucket marker results, where the larger flows gather most of the tokens, we see that the F-SAM marker effectively isolates flows among the aggregates. It also evenly distributes tokens among the individual flows in a max-min fair manner, without maintaining any per-flow state.

## C. End-to-end Performance

In this subsection, we compare the end-to-end performance of PAM and F-SAM with the TB and the sophisticated TSW2CM marker.

In all the experiments in this subsection, we use the topology depicted by Figure 1. The CIR was 1Mb/s, the TB burst size was 500Kb/s and there was no separate background traffic as in the previous set of experiments. All the links except the one between $core$ and $e2$ had a bandwidth of 10Mb/s while the bottleneck had 5Mb/s. The delays in all the links were 5ms. We have used TCP Reno for all TCP agents and the short lived flows were generated by the HTTP 1.0 model in *ns-2*. The parameters for the RIO queues were the default ones in *ns-2*. We have got better results by tweaking the RIO parameters but we have not presented those results here.

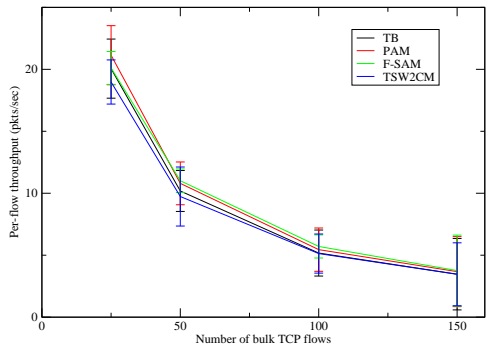

Fig. 4. Average throughput, when only bulk TCP flows are used between $s1$ and $d1$

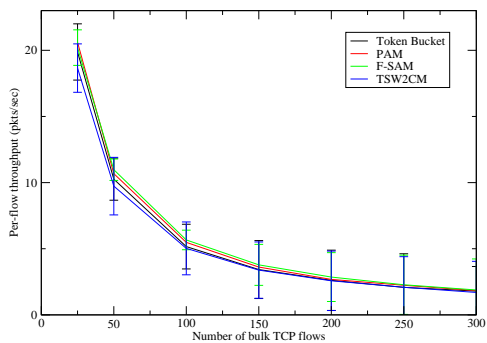

Fig. 5. Average throughput of long flows, when both bulk as well as short lived TCP flows are used between $s1$ and $d1$

First, we tested our markers with long lived bulk TCP flows from $s1$ to $d1$. The results are shown in Figure 4. We see that F-SAM is around 10% better than TB and PAM too performs better than TB (by 2-5%). Both PAM and F-SAM space out the OUT packets probabilistically. Hence, under heavy load, when OUT packets start getting dropped at the congested router, the probability of two packets from the same flow being dropped would become lower. It is interesting to note that both PAM as well as F-SAM have a lesser standard deviation compared to TB, which shows that our markers lead to a more fair bandwidth allocation. The clear winner is F-SAM which has a much lesser standard deviation for the per-flow bandwidth. This is expected as we use a max-min fairness criteria to distribute the token at the marker. Incidentally, TSW2CM performs worse than TB in this scenario.

In the next experiment, we took several bulk TCP flows along with several short lived TCP flows (with an average arrival rate of 1 second and a payload size of 20KB) marked from the same token bucket. We illustrate the results for the long flow and short flows separately. Again, the average per-flow throughput for long flows with PAM and F-SAM yielded higher throughput and less standard deviation (Figure 5). We must note that the F-SAM results have very low standard deviation (almost half of that of the TB marker), which illustrate the fairness of F-SAM marking.

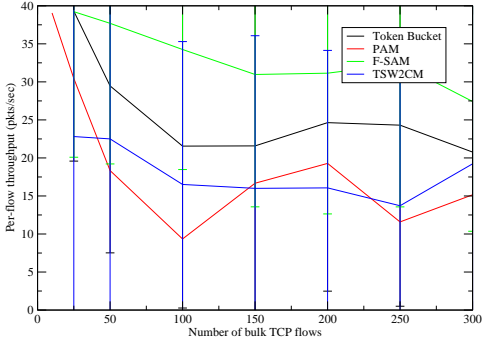

Fig. 6. Average throughput of short flows when both bulk as well as short lived TCP flows are used between $s1$ and $d1$

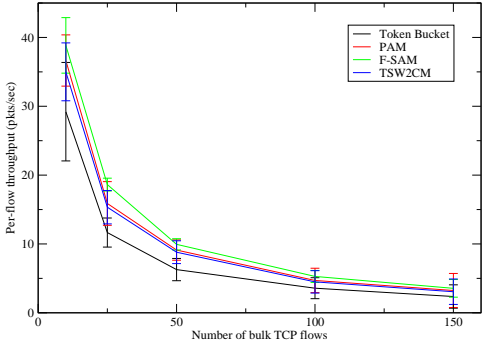

Fig. 7. Average throughput of TCP flows when bulk TCP flows are used between $s1$ and $d1$ and a single misbehaving UDP flow with a rate of 1Mb/s between $s2$ and $d1$

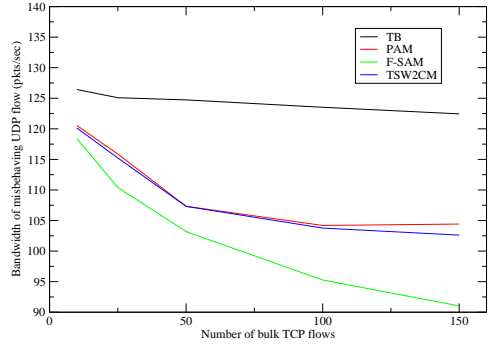

Fig. 8. Bandwidth obtained by the misbehaving flow when bulk TCP flows are present between $s1$ and $d1$ along with a single misbehaving UDP flow with a rate of 1Mb/s between $s2$ and $d1$

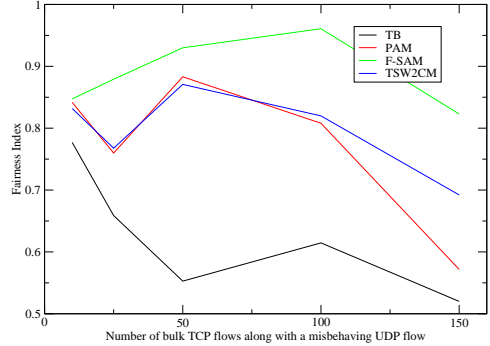

Fig. 9. Comparison of the Fairness Index of markers when bulk TCP flows are present between $s1$ and $d1$ along with a single misbehaving UDP flow with a rate of 1Mb/s between $s2$ and $d1$

Figure 6 depicts the bandwidth obtained by the short flow aggregates. Again, F-SAM beats the competition by a large margin. This is important as we can use F-SAM to win the war between mice and elephants[19] without maintaining per-flow state. Note that we do not need to detect which flows are short as our max-min fairness based F-SAM ensures fair token distribution. Also note that, the bandwidth obtained by mice is almost constant unlike the other markers. Surprisingly, for the short flows, we see that TB is better than PAM.

In the next set of experiments, we used a variable number of bulk TCP flows. Besides, there was a single misbehaving UDP flow generating CBR traffic at 1 Mb/s. The throughput results for the TCP flows are depicted in Figures 7 As a result of our marking, we see that the average bandwidth received by each bulk TCP flows in F-SAM is the highest among all the markers(around 50% more than TB). PAM too is higher than TB by around 30% and is marginally better than TSW2CM. We note that the performance of TSW2CM is closer to PAM since probabilistic rate based marking in TSW2CM too helps to check misbehaving UDP flows. The better TCP performance of F-SAM is clearly due to an evenly distributed out-profile packets compared to bursty out-profile packets in the case of the TB marker in presence of misbehaving UDP flows. This is also demonstrated by Figure 8 shows the bandwidth obtained by the misbehaving UDP flow in the previous scenario. Clearly this result correlates with the previous result (Figure 7, as we see F-SAM (and PAM to a lesser extent) penalizing the UDP flow much more than a TB marker. The fairer token distribution ensures that bandwidth is taken away from the misbehaving flow and is distributed amongst the TCP flows.

Figure 9 plots the fairness index using the throughput of the TCP flows and the UDP flow. The fairness index [21] can be written as

$$FI = \frac{(\sum_i^N x_i)^2}{N \times \sum_i^N x_i^2} \quad (2)$$

where $x_i$ is the number of marked packets per flow or the per-flow throughput. We see a marked increase in the fairness index of F-SAM over all the other markers. PAM too exhibits a much greater fairness index than TB.

The performance improvement for both F-SAM and PAM continues as the bandwidth of the misbehaving flow is increased. We have the complete set of results in [18] This demonstrates the efficacy of our marking schemes. We should note that no special tuning of queue parameters were required to get performance improvement. In fact, we have obtained better performance when we tuned the parameters of the PAM marker as well as the RIO queues.

Thus, these experiments demonstrate that F-SAM does a very good job of isolating misbehaving flows. We conjecture that techniques such as F-SAM will give us the upper bound in aggregate marking. PAM is a much less sophisticated marker and has less dramatic gains. But it does well for all-TCP flows and the implementation overheads are very small. It is easy to see that the two markers provide a different set of benefits. One needs to look at the design requirements and match it with the performance trade-offs when deciding the best marker.

We must note that both our markers are very easily deployable as they do not need any per flow state maintenance. Such simple schemes are promising because their performance shows a marked improvement over simplistic token bucket aggregate markers, and in fact come closer to that of many sophisticated per-flow marking schemes discussed in the related work section.

## VII. Future Work

Each of the the traffic markers introduced needs some more work. We need to explore the design space in more detail. That will ensure it can deployed in a variety of scenarios. We are currently adapting our techniques to segregate the UDP flows early enough so that we can give better fairness guarantees to TCP flows.

For PAM, we plan to incorporate marking schemes based on concepts of PI controllers and try to dynamically auto-configure the RED-like parameters based on the traffic pattern. In F-SAM, we have not yet looked at the idea of weighted fair queuing while distributing tokens among the flows. Allowing for different weights to individual flows of the aggregate is the next logical step after approximating fair queuing. Another issue we have to look at is that the rate estimation being performed at the source leaf node might be incorrect by the time the flow reaches the marker, and this could present some inaccuracies to the fair queuing approximation algorithm.

One direction from this work is to generalize this aggregate marking framework to provide application specific service differentiation. For example, we are investigating how our techniques can be applied to protect short term TCP flows by building a multilevel marker. This can help improve the performance of web traffic. We are currently investigating the use of F-SAM to achieve this. Finally, we want to explore how our markers can be used to implement good, fair pricing schemes apart from designing better architecture for ensuring QoS.

## VIII. Conclusion

In this paper, we have proposed two aggregate markers that ensure fairness amongst different flows in the aggregate without maintaining per-flow state. This makes our approach unique. Probabilistic Aggregate Marking (PAM) ensures fairness in a proportional fashion and allocates tokens to packets with a probability transfer function that looks similar to the transfer function of RED. We also presented a more sophisticated marker called Fair Stateless Aggregate Marker (F-SAM) which is based on fair queueing using *max-min* fairness criteria. The promising aspect of our work is F-SAM which can boost the performance of short lived flows and also ensure fairer bandwidth allocation between TCP and UDP flows.

The above markers are scalable and readily deployable. Our hope is that markers like PAM and F-SAM will enable architectures based on application level marking along with probabilistic aggregate marking. Such architectures promise much more deployable and scalable solution to the problem of traffic marking in the differentiated services framework.

## References

[1] S. Blake, D. Black, M. Carlson, E. Davies, Z. Wang, and W. Weiss, "An architecture for differentiated services," *RFC 2475*, 1998.
[2] K. Nichols, V. Jacobson, and L. Zhang, "A twobit differentiated services architecture for the internet," *RFC 2638*, 1999.
[3] Lixia Zhang, Steve Deering, Deborah Estrin, Scott Shenker, and Daniel Zappala, "Rsvp: A new resource reservation protocol," *IEEE Network Magazine*, September 1993.
[4] J. Heinehan, T. Finner, F. Baker, W. Weiss, and J. Wroclawski, "Assured forwarding phb group," *RFC 2597*, 1999.
[5] B. Braden et al, "Recommendations on queue management and congestion avoidance in the internet," *RFC 2309*, 1998.
[6] S. Floyd and V. Jacobson, "Random early detection gateways for congestion avoidance.," *IEEE/ACM Transactions on Networking, V.1 N.4*, 1993.
[7] R. Pan, B. Prabhakar, and K. Psounis, "A stateless active queue management scheme for approximating fair bandwidth allocation," *IEEE INFOCOM 2000*, 2000.
[8] Ion Stoica, Scott Shenker, and Hui Zhang, "Core-stateless fair queueing: Achieving approximately fair bandwidth allocations in high speed networks," *Sigcomm*, 1998.
[9] A.Feroz, A. Rao, and S. Kalyanaraman, "A tcp-friendly traffic marker for ip differentiated services," *Proc. of the IEEE/IFIP Eighth International Workshop on Quality of Service - IWQoS*, 2000.
[10] J. Heinanen and R. Guerin, "A single rate three color marker," *RFC 2697*, 1999.
[11] J. Heinanen and R. Guerin, "A two rate three color marker," *RFC 2698*, 1999.
[12] W. Fang, N. Seddigh, and B. Nandy, "A time sliding window three colour marker (tswtcm)," 2000.
[13] Ikjun Yeom and A. L. Narasimha Reddy, "Adaptive marking for aggregated flows," *Globecomm*, 2001.
[14] H. Su and Mohammed Atiquzzaman, "Itswtcm: A new aggregate marker to improve fairness in difserv," *Globecomm*, 2001.
[15] Lloyd Wood Ilias Andrikopoulos and George Pavlou, "A fair traffic conditioner for the assured service in a differentiated services internet," *Proceedings of ICC 2000, vol. 2 pp. 806-810*, 2000.
[16] Dong Lin and Robert Morris, "Dynamics of random early detection," *SIGCOMM '97*, pp. 127–137, september 1997.
[17] A. Demers, S. Keshav, and S.J. Shenker, "Analysis and simulation of a fair queueing algorithm," *Sigcomm*, 1989.
[18] A. Das, D. Dutta, and A. Helmy, "Fair stateless aggregate marking techniques for differentiated services networks, university of southern california technical report usc-cs-tr-02-752," 2002.
[19] L. Guo and I. Matta, "The war between mice and elephants," *ICNP*, 2001.
[20] UCB/LBNL/VINT, "The NS2 network simulator, available at http://www.isi.edu/nsnam/ns/.," .
[21] Bruce S. Davie and Larry L. Peterson, *Computer Networks: A Systems Approach*, Morgan Kauffmann, second edition, 2000.